\documentclass[9pt]{article}

\usepackage{spconf,amsmath,graphicx,hyperref}
\hypersetup{hidelinks}
\usepackage{amssymb}
\usepackage{booktabs}
\usepackage{subcaption}
\usepackage{caption} 
\usepackage{array}
\usepackage{tabularx}
\usepackage{array}
\usepackage{multirow}

\title{TRAINING-FREE CONTEXTUAL ASR VIA SPEECHLLM-BASED ERROR-AWARE SELECTIVE RETRIEVAL}

\name{
Natsuo Yamashita$^{1}$,
Ai Nemoto$^{2}$,
Ryosuke Koichi$^{2}$,
Masaaki Yamamoto$^{1}$
}

\address{
$^{1}$Hitachi, Ltd., Research \& Development Group, Japan\\
$^{2}$Hitachi Advanced Systems Corporation, Japan
}

\begin{document}
\ninept
\maketitle

\begin{abstract}
Recognition of domain-specific and low-frequency terms remains challenging for automatic speech recognition (ASR). Although contextual biasing can improve their recognition, directly providing a large terminology dictionary introduces many irrelevant biasing terms. Retrieval-based contextual biasing addresses this issue by selecting candidate terms from an external dictionary, but querying many recognized words requires numerous dictionary lookups and may yield poorly targeted candidates. We propose a training-free contextual ASR framework in which a pretrained speech large language model (SpeechLLM) jointly generates an ASR hypothesis and localizes error spans likely to involve domain-specific terms. Only the localized spans are used to retrieve phonologically similar terms from an external terminology dictionary. The same SpeechLLM then re-recognizes the audio conditioned on the first-pass hypothesis and the retrieved terms, without task-specific model training. To assess applicability across domains, we evaluate the framework on medical, air traffic control, and financial speech. The proposed method substantially reduces dictionary queries while improving the recall and ranking of relevant terminology candidates and second-pass ASR performance across all three domains.
\end{abstract}


\section{Introduction}

Recognition of domain-specific and low-frequency terms remains
challenging for automatic speech recognition (ASR)
~\cite{pundak2018deep,gong2025br}.
Contextual biasing can improve the recognition of such terms by
providing candidate words or phrases at inference time
~\cite{pundak2018deep,sun2021tree,huang2024neural}.
Recent speech large language model (SpeechLLM)-based approaches have
further incorporated pronunciation-aware context and bias-word
information~\cite{fu2025pac,novitasari2026contextual,gong2024contextual}.
However, directly using a large terminology dictionary introduces many
irrelevant candidates, increasing computational cost and potentially
causing over-biasing
~\cite{gong2025br,gong2024contextual,agrawal2025spot}.

To address this issue, prior work has explored several strategies for
selecting terms or regions for retrieval or biasing, including
named-entity-based retrieval~\cite{pusateri2025retrieval}, phonetic
retrieval~\cite{lei2025phonetic}, learned retrieval
~\cite{gong2025br,kong2025glclap}, bias-word ranking and
selection~\cite{hou2025ranking}, terminology relevance
estimation~\cite{du2025attention2probability}, and localized
alignment~\cite{huang2026clar}.
Several recent approaches rely on separately trained retrieval,
ranking, or alignment components
~\cite{gong2025br,kong2025glclap,hou2025ranking,huang2026clar},
introducing additional training and model-development requirements.
Moreover, existing selection criteria do not explicitly target regions
that are likely to correspond to domain terms and contain recognition errors.
This motivates a training-free approach to determining whether terminology retrieval is needed and which regions should trigger it.

In parallel, ASR error detection has been studied to guide selective
correction and decoding
~\cite{lin2023multimodal,he2023edcec,he2025pmfcec}.
However, detected errors have primarily been used to determine where
correction or re-decoding should be applied, rather than whether an
external terminology dictionary should be queried.

\begin{figure}[t]
    \centering
    \setlength{\abovecaptionskip}{2pt}
    \setlength{\belowcaptionskip}{2pt}
    \includegraphics[width=1.0\linewidth]{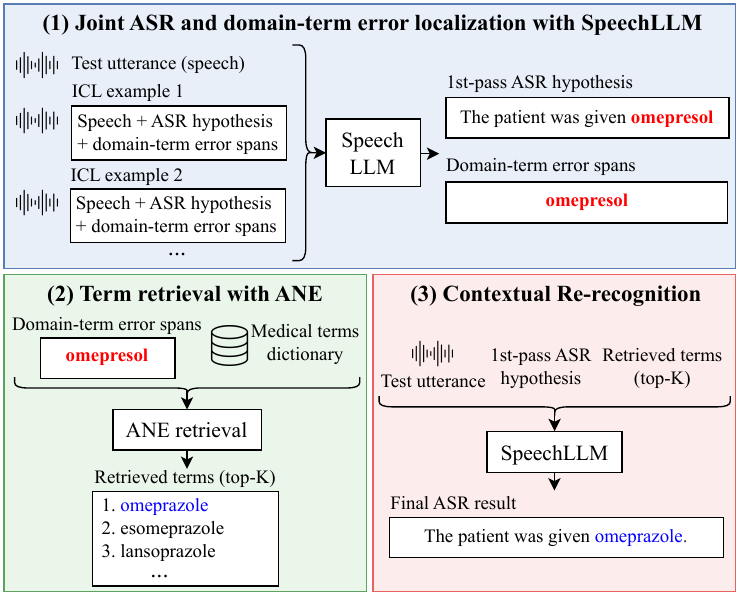}
    \caption{Overview of the proposed training-free contextual ASR framework
with error-aware selective terminology retrieval.}
    \label{fig:proposal}
    \vspace{-10pt}
\end{figure}

SpeechLLMs provide a natural basis for this problem because they can
jointly exploit acoustic information and textual context while
generating an ASR hypothesis~\cite{xu2025qwen3omni}.
If the same pretrained model can also identify domain-term error
regions in its own hypothesis, terminology retrieval can be triggered
without training a separate error-detection model, thereby avoiding
additional task-specific model development.
This motivates the following question: can a pretrained SpeechLLM identify domain-term recognition errors in its
own ASR hypothesis and use them to determine whether terminology retrieval
is needed and which regions should trigger it?

To address this question, we propose a training-free contextual
ASR framework in which a pretrained SpeechLLM jointly generates an ASR hypothesis and localizes error spans likely to involve
domain-specific terms in a single inference.
Rather than directly correcting the localized errors, we use the
predicted spans as a hard gate for terminology retrieval, such that
only regions likely to contain domain-term recognition errors trigger
dictionary search.
Localized spans are used to retrieve phonologically similar domain
terms using Acoustic Neighbor Embeddings (ANE)~\cite{jeon2020ane}.
The same SpeechLLM then re-recognizes the audio conditioned on the
first-pass ASR hypothesis and the retrieved terms.
The entire framework requires no task-specific model training.

Experiments on medical, air traffic control, and financial speech
show that the proposed method substantially reduces dictionary queries while improving the recall and ranking of relevant
terminology candidates and second-pass ASR performance.

\section{Proposed Method}
\label{sec:method}

We propose a training-free contextual ASR framework in which
SpeechLLM-based domain-term error localization is used to selectively
trigger retrieval from an external terminology dictionary.
As illustrated in Fig.~\ref{fig:proposal}, the framework consists of
three steps:
(1) joint generation of a first-pass ASR hypothesis and localization of error spans likely to involve domain-specific terms,
(2) terminology retrieval using only the localized spans as queries,
and
(3) re-recognition of the same audio conditioned on the first-pass ASR
hypothesis and the retrieved terms.
The localized spans serve as a hard gate for dictionary retrieval,
such that only regions predicted to contain domain-term recognition
errors trigger terminology search.
The same pretrained SpeechLLM is used for joint first-pass ASR and
domain-term error localization and for second-pass contextual
re-recognition, with no task-specific parameter updates.

\subsection{Joint ASR and Domain-Term Error Localization}
\label{sec:localization}

Let $\mathbf{x}$ denote an input speech signal, and let
$p_{\mathrm{dom}}$ denote the dataset-specific localization guidance
provided in the prompt.
As exemplified for MedSyn in Table~\ref{tab:prompt}, the guidance
specifies the target domain, relevant terminology categories, and
selection criteria requiring a span to be both domain-relevant and
likely to contain an ASR error.
The same task, selection criteria, ICL format, and output format are
used across datasets, while the domain description, terminology
categories, and demonstrations are adapted to each domain.

Given $\mathbf{x}$ and $p_{\mathrm{dom}}$, the SpeechLLM jointly
generates a first-pass ASR hypothesis
$\mathbf{h}^{\mathrm{1st}}$
and localizes a set of domain-term recognition error spans
$\mathcal{S}=\{s_1,\ldots,s_M\}$.
Each $s_m$ is a short text span extracted from
$\mathbf{h}^{\mathrm{1st}}$ and predicted to contain a recognition
error involving a domain-specific term.
After generation, the predicted spans are normalized and restricted to
at most two words, and numeric-only spans are discarded.

We perform joint ASR and domain-term error localization using
audio-text in-context learning (ICL)~\cite{pan2024cosmic}.
Let
{
\setlength{\abovedisplayskip}{4pt}
\setlength{\belowdisplayskip}{4pt}
\begin{equation}
\mathcal{D}_{\mathrm{ICL}}
=
\{(\mathbf{x}_i,\mathbf{h}_i,\mathcal{S}_i)\}_{i=1}^{J}
\end{equation}
}%
denote the set of $J$ ICL demonstrations, where
$\mathbf{x}_i$ is an example speech signal,
$\mathbf{h}_i$ is its ASR hypothesis, and
$\mathcal{S}_i=\{s_{i,1},\ldots,s_{i,M_i}\}$ is the corresponding
set of domain-term recognition error spans.
We design the ICL demonstrations to cover four
recognition cases: an error-free case, a domain-term substitution,
a domain term misrecognized as a common-word sequence, and a
phonetically similar substitution.
Each demonstration follows the same structured output format as the
test inference, containing an ASR hypothesis followed by its
domain-term error spans.
For each error-containing demonstration, only the ASR-side span
corresponding to the misrecognized domain term is annotated in
$\mathcal{S}_i$, whereas the error-free demonstration uses
$\mathcal{S}_i=\emptyset$.
The demonstrations therefore illustrate both whether a span should be selected and where it should be localized, without providing a corrected transcript as the localization target.

For a test utterance, the structured output is serialized such that the
complete first-pass ASR hypothesis is generated before the domain-term
error spans.
Since autoregressive generation conditions later output tokens on
preceding generated tokens~\cite{xu2025qwen3omni,vaswani2017attention},
this ordering allows span localization to condition directly on the
SpeechLLM's own generated hypothesis.
The joint inference is expressed as
{
\setlength{\abovedisplayskip}{4pt}
\setlength{\belowdisplayskip}{4pt}
\begin{equation}
(\mathbf{h}^{\mathrm{1st}},\mathcal{S})
=
F_{\mathrm{SLLM}}
\left(
\mathbf{x},p_{\mathrm{dom}};
\mathcal{D}_{\mathrm{ICL}}
\right).
\end{equation}
}%
Unlike post-hoc localization applied to a fixed ASR hypothesis, the
proposed method jointly obtains the first-pass hypothesis and its
domain-term recognition error spans within a single SpeechLLM
inference.
The localized spans are not corrected directly at this stage.
Instead, they determine whether terminology retrieval is needed and which regions should trigger it.
Only the predicted spans are used as retrieval queries, while the
complete first-pass hypothesis is retained for subsequent second-pass
re-recognition but is not used to query the terminology dictionary.

\begin{table}[t]
\centering
\setlength{\abovecaptionskip}{2pt}
\setlength{\belowcaptionskip}{2pt}
\caption{Prompt example for joint ASR and domain-term error localization
on MedSyn.}
\label{tab:prompt}
\footnotesize
\begin{tabular}{>{\raggedright\arraybackslash}p{0.95\columnwidth}}
\hline
\noalign{\vskip 3pt}

\textbf{Task:}
Transcribe the input speech and identify short domain-term error spans
in the generated transcript. \\

\textbf{Domain guidance:}
The target domain is medical speech.
Relevant terminology includes medications, diseases, anatomy,
clinical procedures, laboratory tests, and medical abbreviations. \\

\textbf{Selection guidance:}
Select a span only when it is both domain-relevant and likely to be
incorrect.
Do not select a span solely because it contains a domain-specific term.
Use acoustic evidence and the generated transcript to identify
misrecognized, weakly supported, or acoustically unclear regions.
Return short spans corresponding to the ASR-side error expression. \\

\textbf{ICL format:}
Each demonstration consists of an input speech signal and a structured
output containing its ASR hypothesis and domain-term error spans.
The demonstrations include both error-containing and error-free cases. \\

\textbf{Output:}
{\scriptsize\texttt{\{"transcript":\,"...",
"error\_spans":\,[...]\}}} \\

\noalign{\vskip 3pt}
\hline
\end{tabular}
\vspace{-10pt}
\end{table}

\subsection{Terminology Retrieval from Localized Spans}
\label{sec:retrieval}

Each localized span $s_m \in \mathcal{S}$ is used as an independent
retrieval query, yielding
$\mathcal{Q}=\{s_1,\ldots,s_M\}$.
Let $\mathcal{V}=\{v_1,\ldots,v_L\}$ denote an external terminology
dictionary containing $L$ domain-specific terms.
We retrieve terms that are phonetically similar to the localized spans
using Acoustic Neighbor Embeddings (ANE)~\cite{jeon2020ane}.
Let $g(\cdot)$ denote the ANE text encoder.
For a query $q$ and dictionary term $v$, the similarity score is
computed as
{
\setlength{\abovedisplayskip}{3pt}
\setlength{\belowdisplayskip}{3pt}
\begin{equation}
r(q,v)
=
\frac{1}
{1+\left\|g(q)-g(v)\right\|_2}.
\end{equation}
}%

When multiple spans are localized, the similarity scores are aggregated
for each dictionary term as
{
\setlength{\abovedisplayskip}{4pt}
\setlength{\belowdisplayskip}{4pt}
\begin{equation}
R(v \mid \mathcal{Q})
=
\max_{q \in \mathcal{Q}} r(q,v).
\end{equation}
}%
The dictionary terms are ranked by $R(v \mid \mathcal{Q})$ across all
localized spans, and the top-$K$ terms form the candidate set
$\mathcal{C}_K$.
If no span is localized ($M=0$), terminology retrieval and
second-pass re-recognition are skipped, and the first-pass hypothesis
is retained.

\begin{figure*}[t]
  \centering
  \footnotesize

  % ============================================================
  % Table 2
  % ============================================================
  \begin{minipage}[t]{0.42\textwidth}
      \vspace{0pt}
      \centering
      \captionsetup[table]{skip=2pt}
      \captionof{table}{Statistics of the evaluation datasets.}
      \label{tab:dataset}
    
      \renewcommand{\arraystretch}{1}
    
      \begin{tabular}{
        @{}l
        @{\hspace{9pt}}r
        @{\hspace{8pt}}r
        @{\hspace{8pt}}r
        @{\hspace{8pt}}r@{}
      }
        \toprule
        \begin{tabular}{@{}r@{}}
          \\
          Dataset
        \end{tabular}
        &
        \begin{tabular}{@{}r@{}}
          Number of\\
          utterances
        \end{tabular}
        &
        \begin{tabular}{@{}r@{}}
          Reference\\
          words
        \end{tabular}
        &
        \begin{tabular}{@{}r@{}}
          Domain\\
          terms
        \end{tabular}
        &
        \begin{tabular}{@{}r@{}}
          Domain\\
          ratio (\%)
        \end{tabular}
        \\
        \midrule
    
        ATCOSIM~\cite{hofbauer2008atcosim}
        & 1,901 & 22,642 & 167 & 9.00 \\
    
        Earnings~\cite{durmus2026contextualearnings22,delrio2022earnings22}
        & 772 & 26,750 & 652 & 5.04 \\
    
        MedSyn~\cite{united2024medsyn}
        & 7,906 & 106,566 & 7,198 & 11.65 \\
    
        \bottomrule
      \end{tabular}
    \end{minipage}
  \hfill
  % ============================================================
  % Table 3
  % ============================================================
  \begin{minipage}[t]{0.55\textwidth}
  \vspace{0pt}
  \centering
  \captionsetup[table]{skip=2pt}
  \captionof{table}{Audio-text ICL examples used for MedSyn evaluation.}
  \label{tab:icl_examples}

  \renewcommand{\arraystretch}{1}

  \begin{tabular}{
    @{}l
    @{\hspace{24pt}}l
    @{\hspace{12pt}}l
    @{\hspace{12pt}}l@{}
  }
    \toprule
    Error type
    & Reference term
    & ASR hypothesis
    & Error span \\
    \midrule

    Error-free
    & Amoxicillin
    & Amoxicillin
    & \textit{None} \\

    Domain-term substitution
    & Hydralazine
    & Hydroxyzine
    & Hydroxyzine \\

    Common-word substitution
    & Metformin
    & met foreman
    & met foreman \\

    Phonetic substitution
    & ileum
    & ilium
    & ilium \\

    \bottomrule
  \end{tabular}
\end{minipage}
\vspace{-4pt}
\end{figure*}

\subsection{Second-Pass Contextual Re-recognition}
\label{sec:biasing}

The retrieved terms $\mathcal{C}_K$ and the first-pass ASR hypothesis
$\mathbf{h}^{\mathrm{1st}}$ are provided to the same SpeechLLM together
with the input speech for second-pass re-recognition.
The retrieved terms serve as contextual biasing information
~\cite{pundak2018deep,sun2021tree}, while the first-pass hypothesis
provides textual context for re-recognition.
The resulting second-pass ASR hypothesis is expressed as
{
\setlength{\abovedisplayskip}{4pt}
\setlength{\belowdisplayskip}{4pt}
\begin{equation}
\mathbf{h}^{\mathrm{2nd}}
=
F_{\mathrm{SLLM}}
\left(
\mathbf{x},
\mathbf{h}^{\mathrm{1st}},
\mathcal{C}_K
\right).
\end{equation}
}%

% \begin{figure*}[t]
%   \centering
%   \captionsetup[table]{skip=2pt}
%   \captionof{table}{Uncertainty localization performance on MedSyn.}
%   \label{tab:localization}

%   \footnotesize
%   \renewcommand{\arraystretch}{1}

%   \begin{tabular}{
%     l
%     @{\hspace{60pt}}r
%     @{\hspace{10pt}}r
%     @{\hspace{10pt}}r
%     @{\hspace{10pt}}r
%     @{\hspace{10pt}}r
%   }
%     \toprule
%     Method
%     & Spans\,/\,Utt.
%     & Error Precision (\%)
%     & Error Recall (\%)
%     & General Error Recall (\%)
%     & Domain-term Error Recall (\%) \\
%     \midrule

%     \multicolumn{6}{@{}l}{\textbf{Span-selection baselines}} \\

%     All ASR 1-grams
%     & 13.37 & 8.1 & 93.0 & 83.6 & 99.1 \\

%     Random span
%     & 0.99 & 11.5 & 10.6 & 11.0 & 10.3 \\

%     NE extraction
%     & 1.72 & 29.5 & 63.6 & 37.0 & 81.7 \\

%     Confidence span
%     & 1.15 & 49.4 & 49.2 & 28.9 & 65.3 \\

%     \midrule

%     \multicolumn{6}{@{}l}{\textbf{Modality ablations}} \\

%     ASR only
%     & 0.67 & 65.0 & 54.4 & 36.6 & 66.4 \\

%     Audio only
%     & 0.94 & 27.4 & 18.6 & 12.3 & 22.8 \\

%     Audio\,+\,ASR
%     & 1.07 & 49.9 & 67.0 & 41.4 & 84.4 \\

%     \midrule

%     Ours (Joint ICL)
%     & 0.99
%     & 51.9
%     & \textbf{76.1}
%     & \textbf{47.6}
%     & \textbf{86.0} \\

%     \bottomrule
%   \end{tabular}
% \end{figure*}

\begin{figure}[t]
  \centering
  \captionsetup[table]{skip=2pt}
  \captionof{table}{Domain-term error localization performance on MedSyn.
  All metrics except Spans\,/\,Utt. are reported in percent.}
  \label{tab:localization}

  \footnotesize
  \renewcommand{\arraystretch}{1.00}

  \begin{tabular}{
    @{}l
    @{\hspace{33pt}}r
    @{\hspace{8pt}}r
    @{\hspace{8pt}}r
    @{\hspace{8pt}}r
    @{\hspace{8pt}}r@{}
  }
    \toprule
    \begin{tabular}{@{}r@{}}
      \\
      Method
    \end{tabular}
    &
    \begin{tabular}{@{}r@{}}
      Spans\\
      /\,Utt.
    \end{tabular}
    &
    \begin{tabular}{@{}r@{}}
      Error\\
      Precision
    \end{tabular}
    &
    \begin{tabular}{@{}r@{}}
      Error\\
      Recall
    \end{tabular}
    &
    \begin{tabular}{@{}r@{}}
      General\\
      Recall
    \end{tabular}
    &
    \begin{tabular}{@{}r@{}}
      Domain\\
      Recall
    \end{tabular}
    \\
    \midrule

    \multicolumn{6}{@{}l}{\textbf{Span-selection baselines}} \\

    All ASR 1-grams
    & 13.37 & 8.1 & 92.8 & 83.6 & 99.1 \\

    Random span
    & 0.99 & 11.5 & 10.6 & 11.0 & 10.3 \\

    NE extraction
    & 1.72 & 29.5 & 63.6 & 37.0 & 81.7 \\

    Confidence span
    & 1.15 & 49.4 & 50.6 & 28.9 & 65.3 \\

    \midrule

    \multicolumn{6}{@{}l}{\textbf{SpeechLLM input configurations}} \\

    ASR only
    & 0.67 & 65.0 & 54.4 & 36.6 & 66.4 \\

    Audio only
    & 0.94 & 27.4 & 18.6 & 12.3 & 22.8 \\

    Audio\,+\,ASR
    & 1.07 & 49.9 & 67.0 & 41.4 & 84.4 \\

    Joint zero-shot
    & 1.09 & 46.7 & 75.6 & 49.7 & 85.9 \\

    \midrule

    Ours (Joint ICL)
    & 0.99
    & 51.9
    & 76.1
    & 47.6
    & 86.0 \\

    \bottomrule
  \end{tabular}
  \vspace{-6pt}
\end{figure}

\section{Experiments}
\label{sec:experiments}

\subsection{Experimental Setup}
\label{sec:exp_setup}

We evaluate the proposed method on three domain-specific speech
datasets: United-MedSyn~\cite{united2024medsyn} (MedSyn) for medical
speech, ATCOSIM~\cite{hofbauer2008atcosim} for air traffic control,
and Contextual Earnings-22
~\cite{durmus2026contextualearnings22,delrio2022earnings22}
(Earnings) for financial speech.
Owing to the large size of MedSyn, we randomly sample 7,906 utterances,
corresponding to 10\% of the dataset, for all experiments.
Table~\ref{tab:dataset} summarizes the evaluation sets and terminology dictionaries, where the domain ratio
denotes the proportion of reference words corresponding to dictionary terms.
Following prior work~\cite{yamashita2026synthetic,le2021contextualized},
we construct each terminology dictionary from reference transcriptions
by retaining target-domain unigrams occurring at most once per million
words in LibriSpeech~\cite{panayotov2015librispeech},
Common Voice~\cite{ardila2020commonvoice}, and
GigaSpeech~\cite{chen2021gigaspeech}.
Number words, words shorter than three characters, and non-alphabetic
tokens are excluded.
The resulting dictionary is fixed across evaluation utterances, and
utterance-level reference information is not used during retrieval.

We use Qwen3-Omni-30B-A3B-Instruct~\cite{xu2025qwen3omni} as the
SpeechLLM for both joint localization and second-pass re-recognition,
without parameter updates. 
\textit{Joint ICL} uses four audio-text ICL demonstrations,
whereas all other localization methods use no demonstrations.
\textit{Joint zero-shot} uses the same localization instructions as
\textit{Joint ICL} without demonstrations.
Table~\ref{tab:icl_examples} shows the domain-term portions of the examples.
The demonstrations are selected from the training set and are
disjoint from the evaluation utterances.
The reference-term column is shown only for interpretation
and is not provided to the model.
Predicted spans are limited to two words, and numeric-only spans are
excluded from retrieval.

We use the publicly released pretrained ANE model
{\footnotesize\texttt{embedder-\\64}}~\cite{jeon2024theoretical}
without additional training to rank dictionary terms for each predicted span.
We refer to ASR-only hypotheses generated by the same SpeechLLM as
\textit{Baseline ASR}. 
All non-joint methods share these fixed
hypotheses, whereas \textit{Joint zero-shot} and \textit{Joint ICL}
jointly generate their own hypotheses and spans.
Retrieval and second-pass settings are otherwise shared across methods.

\subsection{Domain-Term Error Localization}
\label{sec:localization_results}

We compare \textit{Joint ICL} with four span-selection baselines,
three non-joint SpeechLLM input configurations, and a joint zero-shot
ablation.
\textit{All ASR 1-grams} selects every non-numeric word in the
\textit{Baseline ASR} hypothesis.
\textit{Random span} randomly samples spans while matching the number
and word lengths of the \textit{Joint ICL} spans for each utterance.
\textit{NE extraction}, motivated by named-entity-based query
generation~\cite{pusateri2025retrieval}, prompts the same SpeechLLM
to extract named entities from the \textit{Baseline ASR} hypothesis
without explicitly estimating recognition errors.
\textit{Confidence span}, motivated by conventional ASR confidence
estimation and confidence-aware contextual biasing
~\cite{oneata2021confidence,yang2024confidence}, computes each
word's confidence as the mean log probability of its constituent
decoder tokens and selects words below a dataset-specific threshold.
The threshold is tuned on 500 held-out training utterances for each
dataset, disjoint from the evaluation set, to maximize word-level
ASR error detection F1.
These span-selection methods are also used as query sources in the
subsequent retrieval and contextual ASR experiments.

For the SpeechLLM input configurations, \textit{ASR only} estimates
domain-term error spans from the \textit{Baseline ASR} hypothesis,
\textit{Audio only} estimates them from speech, and
\textit{Audio\,+\,ASR} uses both speech and the \textit{Baseline ASR}
hypothesis.
\textit{Joint zero-shot} jointly generates the first-pass ASR
hypothesis and domain-term error spans within a single SpeechLLM
inference without demonstrations, while \textit{Joint ICL} performs
the same joint generation using audio-text ICL.

Localization is evaluated by the average number of spans per utterance,
Error Precision, Error Recall, General Recall, and Domain Recall.
Recognition errors are identified by minimum-edit-distance
alignment~\cite{qiu2021confidence} and mapped to hypothesis-side
positions.
Deletion errors are mapped to the nearest preceding and
following aligned hypothesis positions when available.
Error Precision is the proportion of hypothesis-side positions covered
by the predicted spans that correspond to recognition errors, whereas
Error Recall is the proportion of recognition-error positions covered
by the predicted spans.
General Recall and Domain Recall evaluate localization of
errors unrelated to and involving dictionary terms, respectively.
For Domain Recall, a domain-term error is considered localized when a
span overlaps its hypothesis-side error location.

Table~\ref{tab:localization} shows the localization results.
\textit{All ASR 1-grams} achieved a Domain Recall of 99.1\%, but
selected 13.37 spans per utterance with an Error Precision of only
8.1\%, providing broad coverage at the cost of poor selectivity.
In contrast, \textit{Joint ICL} selected only 0.99 spans per utterance
while achieving 51.9\% Error Precision, 76.1\% Error Recall, and
86.0\% Domain Recall.
Notably, its Domain Recall was substantially higher than its General
Recall of 47.6\%, indicating that the localized spans were more
sensitive to recognition errors involving domain-specific terms.
This may partly reflect that some ASR errors remain linguistically
plausible in context, making them difficult to detect from textual
cues alone~\cite{gekhman2022red}.
Compared with \textit{NE extraction}, \textit{Joint ICL} reduced the
number of selected spans from 1.72 to 0.99 while improving Error
Precision from 29.5\% to 51.9\% and Domain Recall from 81.7\% to
86.0\%.
Compared with \textit{Confidence span}, it used fewer spans while
improving Error Recall from 50.6\% to 76.1\% and Domain Recall from
65.3\% to 86.0\%.

The ablation results further show that both acoustic and textual
information contribute to localization.
Domain Recall was 22.8\% with \textit{Audio only}, 66.4\% with
\textit{ASR only}, and 84.4\% with \textit{Audio\,+\,ASR}, indicating
that the two modalities provide complementary cues.
\textit{Joint zero-shot} further achieved 75.6\% Error Recall and
85.9\% Domain Recall.
Adding audio-text ICL increased Error Precision from 46.7\% to
51.9\% while maintaining similar Error Recall
(75.6\% vs.\ 76.1\%) and Domain Recall
(85.9\% vs.\ 86.0\%), with fewer selected spans
(1.09 vs.\ 0.99 per utterance).
In addition, \textit{Joint zero-shot} and \textit{Joint ICL} achieved
first-pass WERs of 7.59\% and 7.30\%, respectively, compared with
7.30\% for \textit{Baseline ASR}.
Thus, \textit{Joint ICL} localizes error spans without degrading
transcription accuracy on MedSyn.

\begin{figure}[t]
  \centering
  \begin{minipage}[t]{1.0\linewidth}
    \vspace{0pt}
    \centering
    \captionsetup[table]{skip=2pt}
    \captionof{table}{ANE retrieval performance on MedSyn.
    All metrics except Queries\,/\,Utt. are reported in percent.}
    \label{tab:retrieval}

    \footnotesize
    \renewcommand{\arraystretch}{1.0}

    \begin{tabular}{
      @{}l
      @{\hspace{7pt}}r
      @{\hspace{7pt}}r
      @{\hspace{7pt}}r
      @{\hspace{7pt}}r
      @{\hspace{7pt}}r
      @{\hspace{7pt}}r@{}
    }
      \toprule

      \begin{tabular}{@{}l@{}}
        \\
        Method
      \end{tabular}
      &
      \begin{tabular}{@{}r@{}}
        Queries\\
        /\,Utt.
      \end{tabular}
      &
      \begin{tabular}{@{}r@{}}
        \\
        Reduction
      \end{tabular}
      &
      \begin{tabular}{@{}r@{}}
        \\
        R\hspace{-0.05em}@\hspace{-0.05em}1
      \end{tabular}
      &
      \begin{tabular}{@{}r@{}}
        \\
        R\hspace{-0.05em}@\hspace{-0.05em}10
      \end{tabular}
      &
      \begin{tabular}{@{}r@{}}
        \\
        R\hspace{-0.05em}@\hspace{-0.05em}50
      \end{tabular}
      &
      \begin{tabular}{@{}r@{}}
        \\
        MRR\hspace{-0.05em}@\hspace{-0.05em}10
      \end{tabular}
      \\
      \midrule

      All ASR 1-grams
      & 13.37 & 0.0 & 20.7 & 53.2 & 63.7 & 31.9 \\

      Random span
      & 0.99 & 92.6 & 3.7 & 6.1 & 8.4 & 4.4 \\

      NE extraction
      & 1.72 & 87.1 & 30.9 & 52.6 & 64.4 & 38.1 \\

      Confidence span
      & 1.15 & 91.4 & 31.3 & 45.4 & 54.3 & 35.8 \\

      \midrule

      Ours (Joint ICL)
      & 0.99
      & 92.6
      & \textbf{41.4}
      & \textbf{61.8}
      & \textbf{73.1}
      & \textbf{47.9} \\

      \bottomrule
    \end{tabular}
  \end{minipage}%
  \vspace{-0pt}
\end{figure}

\begin{figure}[t]
  \centering
  \begin{minipage}[t]{1.0\linewidth}
    \vspace{0pt}
    \captionsetup[table]{skip=2pt}
    \captionof{table}{Second-pass ASR performance on MedSyn (\%).}
    \label{tab:asr}

    \footnotesize
    \renewcommand{\arraystretch}{1.0}

    \centering
    \begin{tabular}{
      @{}
      l
      @{\hspace{16pt}}r
      @{\hspace{5pt}}r
      @{\hspace{5pt}}r
      @{\hspace{12pt}}r
      @{\hspace{5pt}}r
      @{\hspace{5pt}}r
      @{}
    }
      \toprule
        & \multicolumn{3}{c}{Top-10}
        & \multicolumn{3}{c}{Top-50} \\
        \cmidrule(r{11pt}){2-4}
        \cmidrule(l{0pt}){5-7}
        
        Method
        & WER & B-WER & U-WER
        & WER & B-WER & U-WER \\
      \midrule

      Baseline ASR
      & 7.30 & 41.43 & 2.80
      & 7.30 & 41.43 & 2.80 \\

      All ASR 1-grams
      & 5.83 & 28.83 & 2.80
      & 5.99 & 30.42 & 2.77 \\

      Random span
      & 7.12 & 39.95 & 2.79
      & 7.09 & 39.81 & 2.78 \\

      NE extraction
      & 5.89 & 29.71 & 2.75
      & 6.01 & 30.80 & 2.74 \\

      Confidence span
      & 6.08 & 31.40 & 2.75
      & 6.22 & 32.52 & 2.75 \\

      \midrule

      Ours (Joint ICL)
      & \textbf{5.67} & \textbf{28.05} & \textbf{2.72}
      & \textbf{5.88} & \textbf{29.79} & \textbf{2.72} \\

      \bottomrule
    \end{tabular}
  \end{minipage}
  \vspace{-4pt}
\end{figure}

\subsection{Terminology Retrieval and Contextual ASR}
\label{sec:retrieval_asr_results}

We first evaluate the effectiveness of the localized spans as
terminology retrieval queries.
Retrieval performance is measured by the average number of queries
per utterance, query reduction relative to \textit{All ASR 1-grams},
R\hspace{-0.05em}@\hspace{-0.05em}$K$, and MRR\hspace{-0.05em}@\hspace{-0.05em}10.
R\hspace{-0.05em}@\hspace{-0.05em}$K$ denotes recall of domain terms
involved in errors in each method's first-pass hypothesis within the
top-$K$ candidates, while MRR\hspace{-0.05em}@\hspace{-0.05em}10 is
their mean reciprocal rank, with misses in the top 10 assigned zero.

Table~\ref{tab:retrieval} shows the ANE retrieval results.
\textit{Joint ICL} reduced the number of queries from 13.37 to 0.99
per utterance, corresponding to a 92.6\% reduction relative to
\textit{All ASR 1-grams}, while improving \mbox{R\hspace{-0.05em}@\hspace{-0.05em}10} from 53.2\% to
61.8\% and MRR\hspace{-0.05em}@\hspace{-0.05em}10 from 31.9\% to 47.9\%.
In contrast, \textit{Random span}, which used the same number of
queries as \textit{Joint ICL}, achieved only 6.1\% R\hspace{-0.05em}@\hspace{-0.05em}10.
\textit{Joint ICL} also achieved higher R@K and MRR@10 than
\textit{NE extraction} and \textit{Confidence span}.

We next evaluate second-pass ASR using the retrieved terms together
with the first-pass ASR hypothesis.
Following prior work on contextual ASR~\cite{le2021contextualized},
we report overall WER, B-WER for dictionary terms, and U-WER for
non-dictionary terms using the top-10 and top-50 retrieved candidates.
Table~\ref{tab:asr} shows the second-pass ASR results.
Compared with \textit{Baseline ASR}, \textit{Joint ICL} with the
top-10 candidates reduced WER from 7.30\% to 5.67\% and B-WER from
41.43\% to 28.05\%, while reducing U-WER from 2.80\% to 2.72\%.
Despite using substantially fewer retrieval queries, \textit{Joint ICL}
also achieved lower WER and B-WER than \textit{All ASR 1-grams}
(5.67\% vs. 5.83\% and 28.05\% vs. 28.83\%, respectively).
\textit{Random span} performed considerably worse despite using the
same number of queries as \textit{Joint ICL}.
For \textit{Joint ICL}, the top-10 condition outperformed the top-50
condition, suggesting that additional lower-ranked candidates can
introduce less relevant biasing terms.

\begin{figure}[t]
  \centering
  \begin{minipage}[t]{1.0\linewidth}
    \vspace{0pt}
    \centering
    \captionsetup[table]{skip=2pt}
    \captionof{table}{Cross-domain ANE retrieval and second-pass ASR
performance with top-10 retrieved candidates. All metrics except Queries\,/\,Utt. are reported in percent.}
    \label{tab:cross_domain}

    \footnotesize
    \renewcommand{\arraystretch}{1.0}

    \begin{tabular}{
      @{}l
      @{\hspace{7pt}}l
      @{\hspace{4pt}}r
      @{\hspace{5pt}}r
      @{\hspace{5pt}}r
      @{\hspace{5pt}}r
      @{\hspace{5pt}}r@{}
    }
      \toprule

      Dataset
      &
      Method
      &
      \begin{tabular}[b]{@{}r@{}}
        Queries\\
        /\,Utt.
      \end{tabular}
      &
      R\hspace{-0.05em}@\hspace{-0.05em}10
      &
      MRR\hspace{-0.05em}@\hspace{-0.05em}10
      &
      WER
      &
      B-WER
      \\
      \midrule

      \multirow{6}{*}{ATCOSIM}
      & Baseline ASR
      &
      &
      &
      &
      14.81
      & 46.40 \\

      & All ASR 1-grams
      & 5.89
      & 66.6
      & 34.5
      & 13.13
      & 32.42 \\

      & Random span
      & 1.08
      & 28.1
      & 17.1
      & 14.08
      & 41.00 \\

      & NE extraction
      & 1.51
      & 61.5
      & 37.5
      & 13.06
      & 31.97 \\

      & Confidence span
      & 0.97
      & 34.3
      & 22.9
      & 13.68
      & 36.93 \\
      \cmidrule{2-7}

      & Ours (Joint ICL)
      & 1.08
      & \textbf{69.6}
      & \textbf{46.1}
      & \textbf{12.88}
      & \textbf{30.65} \\

      \midrule

      \multirow{6}{*}{Earnings}
      & Baseline ASR
      &
      &
      &
      &
      17.97
      & 50.93 \\

      & All ASR 1-grams
      & 32.71
      & 31.8
      & 14.8
      & 17.68
      & 47.15 \\

      & Random span
      & 1.38
      & 5.6
      & 3.6
      & 17.80
      & 49.74 \\

      & NE extraction
      & 2.26
      & 35.6
      & 22.2
      & 17.55
      & 46.32 \\

      & Confidence span
      & 1.22
      & 17.6
      & 12.3
      & 17.71
      & 46.85 \\
      \cmidrule{2-7}

      & Ours (Joint ICL)
      & 1.38
      & \textbf{35.8}
      & \textbf{22.5}
      & \textbf{17.45}
      & \textbf{45.66} \\

      \bottomrule
    \end{tabular}
  \end{minipage}%
  \vspace{-4pt}
\end{figure}

\subsection{Cross-Domain Validation}
\label{sec:cross_domain}

We evaluate the proposed method on ATCOSIM and Earnings to examine
its effectiveness beyond the medical domain.
Table~\ref{tab:cross_domain} reports terminology retrieval and
second-pass ASR performance.
Across both datasets, \textit{Joint ICL} substantially reduced dictionary
queries while achieving the highest R\hspace{-0.05em}@\hspace{-0.05em}10 and MRR\hspace{-0.05em}@\hspace{-0.05em}10 among the
evaluated query-selection methods. On Earnings, queries decreased
from 32.71 to 1.38 per utterance while R\hspace{-0.05em}@\hspace{-0.05em}10 increased from 31.8\%
to 35.8\%. 
Random span, which used a comparable number of queries,
performed substantially worse than \textit{Joint ICL}.

\textit{Joint ICL} achieved first-pass WERs of 14.91\% and 17.97\%
on ATCOSIM and Earnings, compared with 14.81\% and 17.97\% for
\textit{Baseline ASR}.
Together with the MedSyn result, this indicates that joint localization does not substantially degrade first-pass transcription accuracy across domains.

For second-pass ASR, \textit{Joint ICL} achieved the lowest WER and
B-WER among the evaluated methods on both datasets. 
Compared with \textit{Baseline ASR}, B-WER decreased from 46.40\% to 30.65\% on
ATCOSIM and from 50.93\% to 45.66\% on Earnings. 
The smaller overall
WER gain on Earnings may reflect its lower domain-term ratio.
These results show that the proposed selective retrieval strategy is
effective across medical, air traffic control, and financial speech.

\section{Conclusion}
\label{sec:conclusion}

We proposed a training-free contextual ASR framework using
SpeechLLM-based domain-term error localization to selectively trigger
terminology retrieval.
The same SpeechLLM performs second-pass re-recognition using the
first-pass ASR hypothesis and retrieved terms.
Cross-domain experiments showed that the proposed method substantially
reduces dictionary queries while improving the recall and
ranking of relevant terminology candidates and second-pass ASR performance.
\vfill\pagebreak

\bibliographystyle{IEEEbib}
\bibliography{strings,refs}

\end{document}